\documentclass{desyproc}

\newcommand{\ljets}   {$\ell$+jets}
\newcommand{\ifb}     {fb$^{-1}$}
\newcommand{\ttbar}   {\mbox{$t\bar{t}$}}
\newcommand{\pt}      {\ensuremath{p_T}}
\newcommand{\mt}      {\mbox{\ensuremath{m_t}}}
\newcommand{\mtmsbar} {\mbox{\ensuremath{m_t^{\overline{\rm MS}}}}}
\newcommand{\mtpole}  {\mbox{\ensuremath{m_t^{\rm pole}}}}
\newcommand{\mtmc}    {\mbox{\ensuremath{m_t^{\rm MC}}}}
\newcommand{\msbar}   {\mbox{$\overline{\rm MS}$}}
\newcommand{\mtt}     {\ensuremath{M_{\mathrm{T}2}}}
\newcommand{\mttp}    {\ensuremath{M_{\mathrm{T}2\perp}}}
\newcommand{\PTm}     {\ensuremath{{p}_\mathrm{T}\hspace{-0.9em}/\kern 0.5em}}
\newcommand{\vPTm}    {\vec\PTm}
\usepackage{multirow}
\begin{document}

\title{Alternative methods for top-quark mass determination at the Tevatron and LHC}
\author{{\slshape Fr\'ed\'eric D\'eliot}  for the ATLAS, CDF, CMS and D0 Collaborations\\[1ex]
CEA-Saclay, DSM/Irfu/SPP, 91191 Gif-sur-Yvette Cedex, France}

\contribID{xy}  
\confID{7095}
\desyproc{DESY-PROC-2013-XY}
\acronym{TOP2013}
\doi            

\maketitle


\begin{abstract}
  I am summarizing here the determination of the top-quark mass from the 
  CDF and D0 Collaborations at the Tevatron and the ATLAS and CMS Collaborations
  at the LHC using non-conventional methods.
  I am concentrating on the extraction of the top-quark mass from the top-antitop cross-section,
  on the mass measurement using the so-called endpoint method and on the top-quark mass
  determination from the $b$-lifetime.
  Other alternative methods are described in another article~\cite{fuster}.

\end{abstract}

\section{Motivations}
\label{sec:motivations}

The top-quark mass is now measured with a remarkable precision around 0.5~\% both
at the Tevatron and at the LHC using well-developped ``standard'' methods based on templates,
matrix elements or ideogram~\cite{velev,brock}.
Despite this precision, some questions remain. Indeed since the top quark is a color object,
it is non trivial to know which mass is really measured using these standard methods.
In all standard methods, Monte Carlo (MC) is used to calibrate the measurements.
This mass implemented in MC generators is different from a well-defined mass in theory.
A way to get some hints about these points experimentally is to determine the top-quark mass
using alternative methods. Such methods can use less inputs from MC or can have different
sensitivity to systematic uncertainties than the standard analyses.
In this article I will concentrate on the extraction of the top-quark mass from the top-antitop (\ttbar) 
cross-section, on the mass measurement using the so-called endpoint method and on the top-quark mass
determination from the $b$-lifetime.
Other alternative methods are described in another article~\cite{fuster}.

Before studying methods which rely differently on MC, it is interesting to look at
the dependence of the measured top-quark mass using standard methods with the event kinematics
and to compare the data measurements with the predictions from MC.
This allows to test the description of the top-quark mass by MC in various phase space regions
and to detect potential large deviations due to the pole mass definition problem described above.
This has been looked at by the CMS Collaboration~\cite{cms} in the \ljets\ final state asking 
for two $b$-tag jets 
using 5~\ifb\ of LHC at 7~TeV~\cite{cms:massdep}. For these comparisons, the top-antitop final state is fully reconstructed 
and the top-quark mass is measured using the ideogram technique either solely or together with the 
jet energy scale. 
The measurements are compared to Madgraph~\cite{madgraph} with different
Pythia~\cite{pythia} tunes and to MC@NLO~\cite{mcatnlo}.
Differential measurements as a function of several variables have been 
performed~\cite{cms:massdep} that are sensitive to different physics effects.
For instance, the top-quark mass distribution as a function of the opening angles between the two light jets 
(see Figure~\ref{fig:massdep}) or as a function of the pseudo-rapidity ($\eta$) of the hadronic decaying top is sensitive to color reconnection.
The influence of initial and final state radiation can be investigated by looking at the top-quark mass
as a function of the invariant mass of the \ttbar\ pair or as a function of the transverse momentum 
(\pt) of the \ttbar\ pair. 
To test the sensitivity to the $b$-quark kinematics, the top-quark mass is measured as a function
of the transverse momentum or the pseudo-rapidity of the $b$-jet assigned to the hadronic decaying top quark. 
The mass distribution as a function of the distance between the $b$- and $\bar{b}$-jets 
($\Delta R_{b\bar{b}} = \sqrt{\Delta \eta^2 + \Delta \phi^2}$) is also scrutinized (see Figure~\ref{fig:massdep}).
Even if the statistical error on these differential measurements is still large, there is 
currently no indication of specific biases due to the choice of generators.  

\begin{figure}[hb]
\centerline{
\includegraphics[width=0.5\textwidth,height=5.5cm]{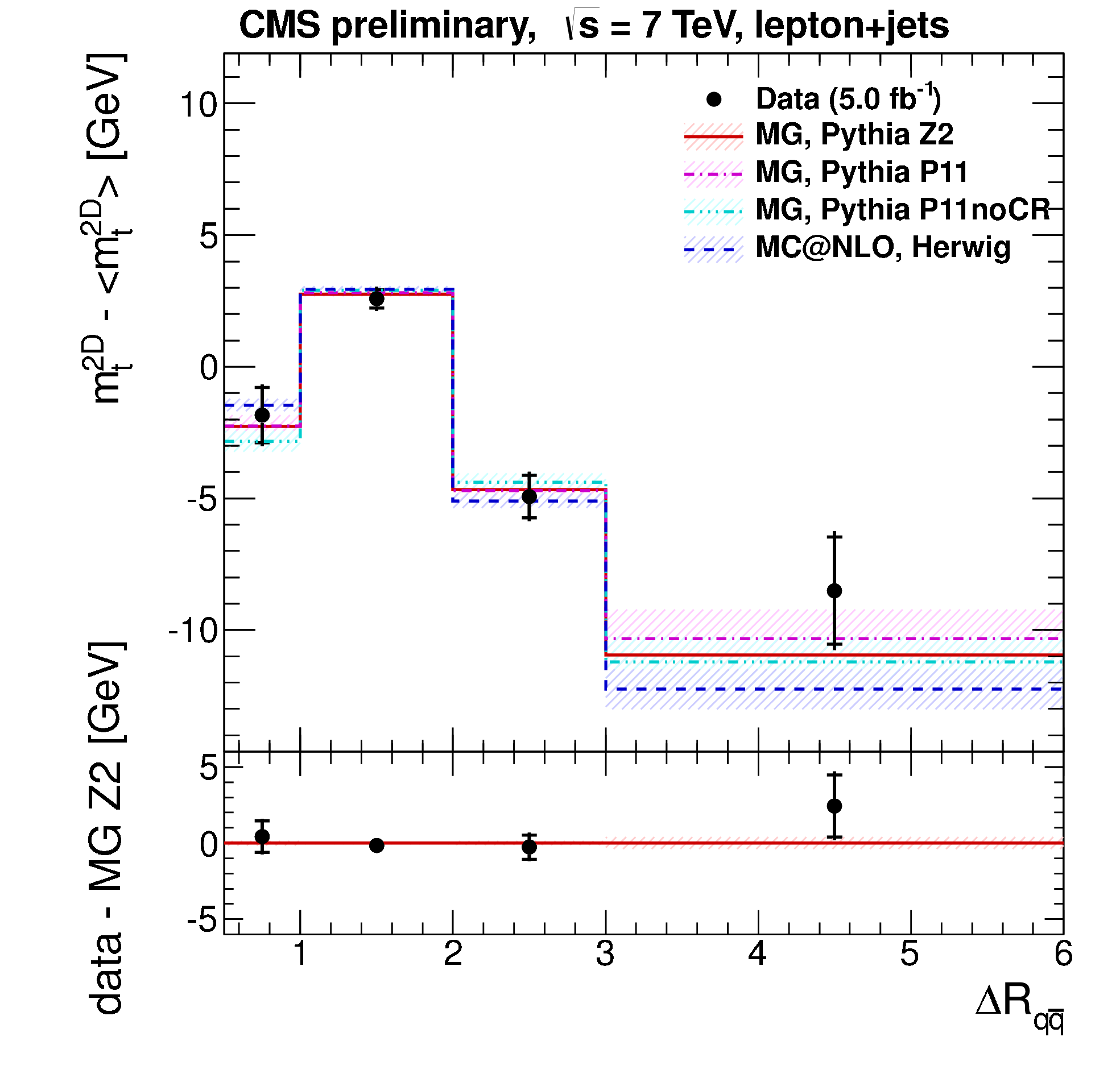}
\includegraphics[width=0.5\textwidth,height=5.5cm]{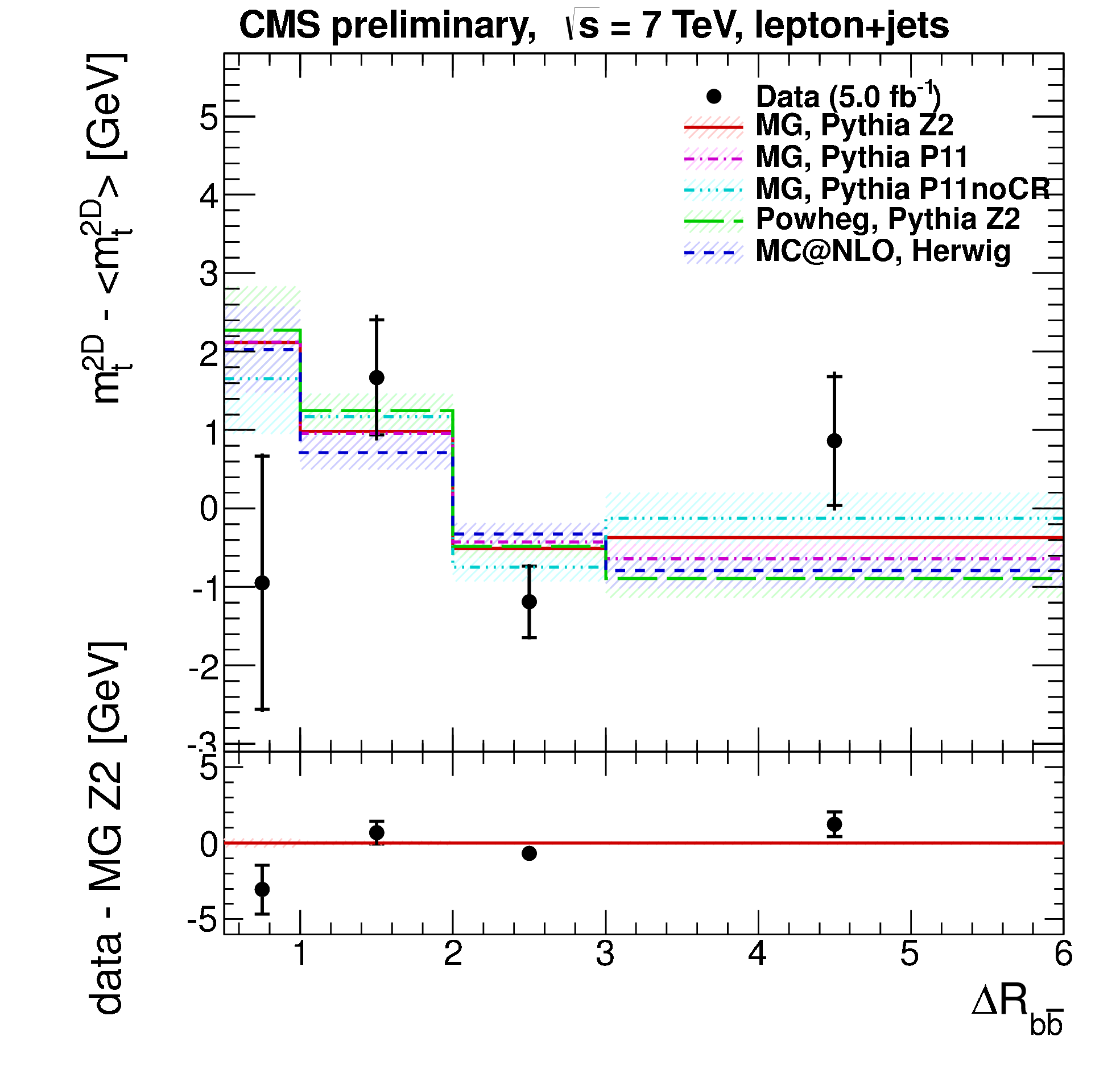}
}
\caption{Differential top-quark mass measurements as a function of the separation of the light-quark jets (left)
and of the b-quark jets (right) performed by CMS~\cite{cms:massdep} compared to several MC predictions.}
\label{fig:massdep}
\end{figure}

\section{Mass extraction from the \ttbar\ cross section}
\label{sec:fromxs}

The principle for the mass extraction from the \ttbar\ cross section is to compare the experimental measured
\ttbar\ cross section with the one computed theoretically. Both the experimental and theoretical cross sections
depend on the top-quark mass but the dependence is different in the two cases. In the experimental case,
the dependency comes from the acceptance cuts while in the theoretical case, it originates from the matrix element.
The advantage of this alternative method lies in the fact that it allows to extract a top-quark mass in a 
well-defined renormalization scheme (the one that in used in the theory computation) in contrast to the one
that is implemented in the MC generators.
This method has however the drawback that it is less precise than direct measurements.


This determination of the top-quark mass has been performed by the D0 Collaboration
using the \ttbar\ cross section measured in the \ljets\ channel using $b$-tagging requirement with
5.4~\ifb. This measured cross section is the one that exhibits the weakest dependence on the top-quark
mass. The variation of the measurement as a function of the MC mass (\mtmc) is parametrized using a third-order
polynomial divided by the mass to the fourth power.
As theory input cross section, the next-to-leading order (NLO), the NLO including next-to-leading log
(NLL) resummation computations and some approximation of the next-to-next-to-leading order (NNLO) calculations.
The mass is extracted from the maximum of a normalized likelihood distribution defined as:
$$L(\mt) = \int f_{\mathrm{exp}} (\sigma | \mt) \, \left[ f_{\rm scale}
(\sigma | \mt) \otimes f_{\rm PDF} (\sigma | \mt) \right] \, d\sigma,$$
where $f_{\mathrm{exp}}$ comes from the experimental measurement which uncertainties are assumed to be Gaussian
distributed, $f_{\rm scale}$ represents the theoretical scale uncertainty, taken to be flat 
and $f_{\rm PDF}$ represents the uncertainty of parton density functions (PDF) taken to be a
Gaussian function.
The mass determination is peformed assuming that \mtmc\ corresponds to the pole mass (\mtpole) and
assuming that \mtmc\ corresponds to the \msbar\ mass (\mtmsbar).
The experimental and theoretical \ttbar\ cross sections used in the extraction are shown in 
Figure~\ref{fig:d0fromxs}. With this technique, D0 measures the top-quark pole mass shown in
Table~\ref{tab:d0fromxs}~\cite{d0:fromxs}. These values are compatible but slightly lower
than the top-quark mass world average~\cite{pdg}. The \msbar\ mass is also extracted~\cite{d0:fromxs}.

\begin{figure}[!htb]
\centerline{
\includegraphics[width=0.45\textwidth,height=5.5cm]{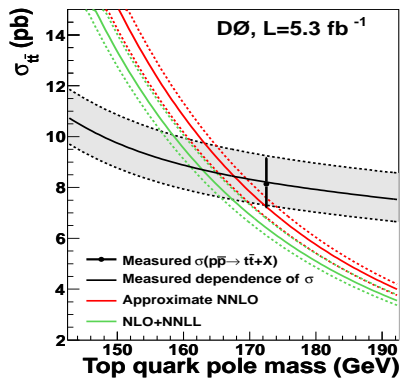}
}
\caption{Experimental and theoretical \ttbar\ cross sections used by D0 to extract the top-quark 
mass~\cite{d0:fromxs}.}
\label{fig:d0fromxs}
\end{figure}

\begin{table}
\centerline{
\begin{tabular}{c|cc}
\hline \hline \\[-9pt]
Theoretical prediction          & {$m_{t}^{\rm pole}$ (GeV)} &
{$\Delta m_{t}^{\rm pole}$ (GeV)} \\[2pt] \hline \\[-8pt]
MC mass assumption  & $m_{t}^{\rm MC} = m_{t}^{\rm pole}$ & $m_{t}^{\rm MC} = m_{t}^{\overline{\rm MS}}$ \\[2pt] \hline \\[-8pt]
NLO               & $164.8^{+5.7}_{-5.4}$ & $-3.0$ \\[2pt]
NLO+NLL           & $166.5^{+5.5}_{-4.8}$ & $-2.7$ \\[2pt]
NLO+NNLL          & $163.0^{+5.1}_{-4.6}$ & $-3.3$ \\ [2pt]
Approximate NNLO  & $167.5^{+5.2}_{-4.7}$ & $-2.7$ \\ [2pt]
\hline \hline
\end{tabular}
}
\caption{Values of the pole top-quark mass \mtpole, with their 68\% C.L. uncertainties extracted 
for different theoretical predictions by D0~\cite{d0:fromxs}.}
\label{tab:d0fromxs}
\end{table}


A similar method has been developped by CMS.
In that analysis, CMS uses the \ttbar\ cross section measured in the dilepton channel using
2.3~\ifb\ at 7~TeV as experimental input. This cross section is the most precise one measured 
by CMS with a total uncertainty of 4.1~\%. As for D0, it is parametrized 
using a third-order polynomial divided by the mass to the fourth power.
The full NNLO prediction including next-to-next-to-leading log (NNLL) resummation 
is employed as theoretical input.
The mass is extracted using a probability function similar to the D0 analysis.
A 1~GeV addition uncertainty is added to the experimental result to cover the possible
difference between \mtmc\ and \mtpole. CMS also studies the interplay of the mass extraction with the value
of the strong coupling constant $\alpha_S$ (see~\cite{naumann} for more details).

\begin{figure}[!htb]
\centerline{
\includegraphics[width=0.5\textwidth]{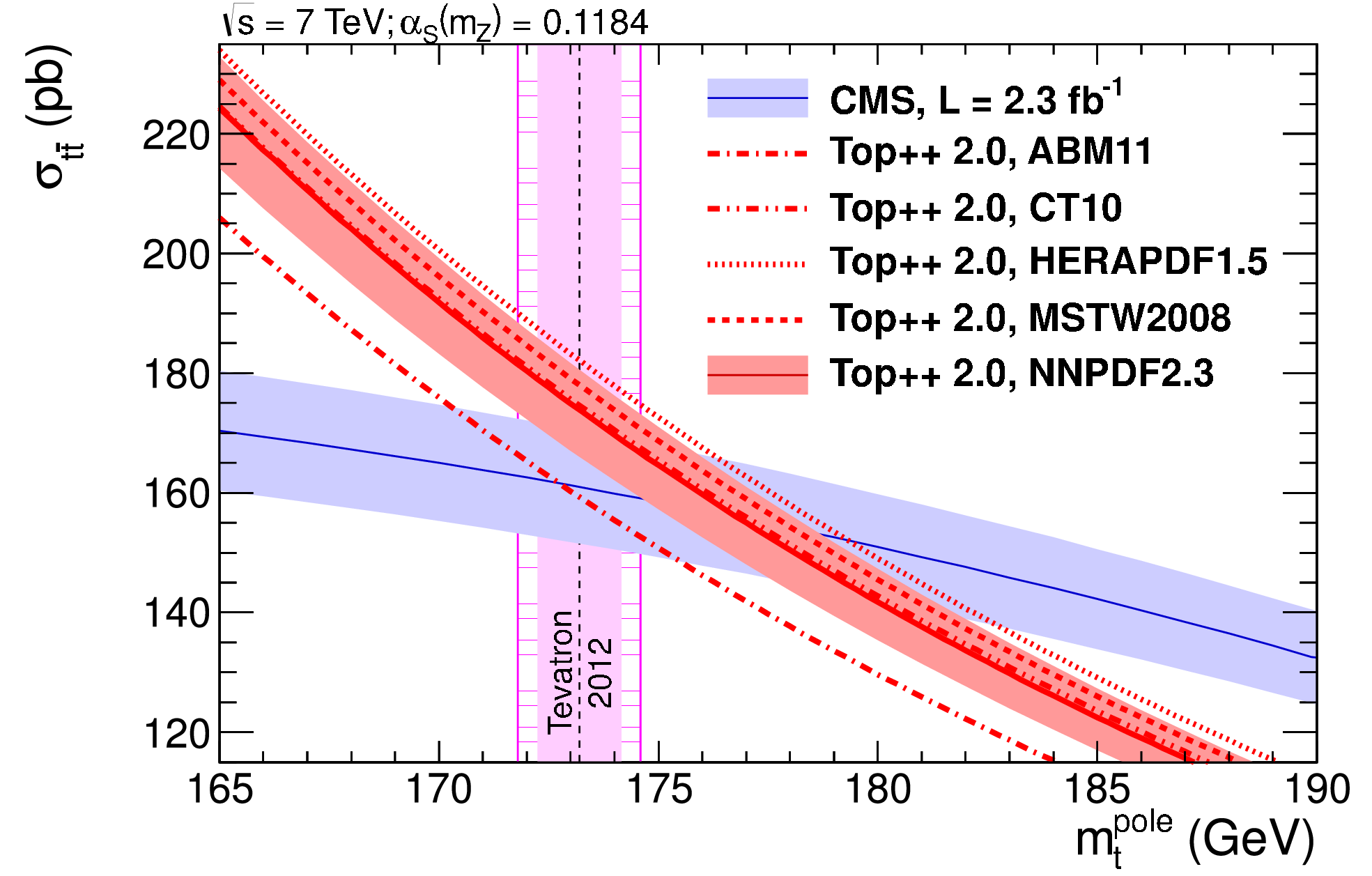}
}
\caption{Experimental and theoretical \ttbar\ cross sections used by CMS to extract the top-quark
mass~\cite{cms:fromxs}.}
\label{fig:cmsfromxs}
\end{figure}

The experimental and theoretical \ttbar\ cross sections used in the extraction are shown in
Figure~\ref{fig:cmsfromxs}. The extracted top-quark pole mass by CMS are shown in
Table~\ref{tab:cmsfromxs} for different PDFs~\cite{cms:fromxs}. These values are compatible but slightly higher
than the top-quark mass world average~\cite{pdg}.
The same kind of extraction has been also performed by the ATLAS Collaboration~\cite{atlas} using 
the first 35~pb$^{-1}$
of LHC data leading to a top-quark mass of $\mtpole = 166^{+7.8}_{-7.3}$~GeV~\cite{atlas:fromxs}.

\begin{table}
\centerline{
\begin{tabular}{l|c|ccc}
\hline
 & Most likely $\mtpole$ & \multicolumn{3}{c}{Uncertainty (GeV)} \\ \cline{3-5}
 & value (GeV)        & Total & From $\delta \alpha_{S}$
 & From $\delta E_{\text{LHC}}$ \\
\hline
ABM11 &  172.7 & ${}^{+3.8}_{-3.5}$ & ${}^{+1.0}_{-1.0}$ & ${}^{+0.8}_{-0.8}$\\
CT10 &  177.0 & ${}^{+4.3}_{-3.8}$ & ${}^{+0.8}_{-0.8}$ & ${}^{+0.9}_{-0.9}$\\
HERAPDF1.5 &  179.5 & ${}^{+4.3}_{-3.8}$ & ${}^{+1.2}_{-1.1}$ & ${}^{+1.0}_{-1.0}$\\
MSTW2008 &  177.9 & ${}^{+4.0}_{-3.6}$ & ${}^{+0.9}_{-0.9}$ & ${}^{+0.9}_{-0.9}$\\
NNPDF2.3 &  176.7 & ${}^{+3.8}_{-3.4}$ & ${}^{+0.7}_{-0.7}$ & ${}^{+0.9}_{-0.9}$\\
\hline
\end{tabular}
}
\caption{Results obtained by CMS for $\mtpole$ by comparing the measured \ttbar\ cross section to
      the NNLO+NNLL prediction with different NNLO PDF sets~\cite{cms:fromxs}.
}
\label{tab:cmsfromxs}
\end{table}


\quad \\

To summarize, the top-quark pole mass has been extracted from the \ttbar\ cross section 
by D0 leading to a precision of 3~\% (where the input experimental cross section has a precision
of 12~\%, and the input theoretical cross section of 3~\%), by ATLAS with a precision of 
4.5~\% (where the experimental input has an uncertainty of 13~\% and the theory input of 5~\%)
and by CMS with a precision of 2~\% (where the experimental input has an uncertainty of 4~\% and the 
theory input of 4~\%). Looking at the current theoretical uncertainty on the \ttbar\ cross section
and assuming no experimental errors, one can estimate the ultimate uncertainty on the top-quark mass 
achievable with this method to be around 3~GeV (1.7 \%).

\section{Mass measurement using the endpoint method}
\label{sec:endpoint}

The endpoint method employed for the first time by CMS to measure the top-quark mass~\cite{cms:endpoint}
was originally developed to measure masses of potentially pair produced new particles
with two cascade decays each ending in an invisible particle, like neutralino. It is thus also applicable
to the \ttbar\ dilepton final state which contains two escaping neutrinos. 
This method relies on the end distribution of the variable named \mtt\ used as mass estimator.
This $M_{T2}$ is a generalization of the usual transverse mass and is defined as:
$$\mtt \equiv
        \min_{\vec{p}^{\text{a}}_{\text{T}}+\vec{p}^{\text{b}}_{\text{T}}=\vPTm}\;
        \left\{\max(M^{\text{a}}_{\text{T}},M^{\text{b}}_{\text{T}})\right\}.$$
This variable corresponds to the minimum parent mass consistent with the observed kinematics
for the hypothetical $\vec{p}^{\text{a}}_{\text{T}}$ and $\vec{p}^{\text{b}}_{\text{T}}$. 
To limit the sensitivity to the transverse momentum of the \ttbar\ system ($\pt(\ttbar)$), the variable \mttp\
is rather used. It is computed with the \pt\ components perpendicular to $\pt(\ttbar)$.

Three variables are needed to solve the dilepton event kinematics.
The chosen variables are \mttp\ computed at the lepton level ($\mu_{\ell\ell}$) after the $W$-boson decays, 
\mttp\ computed at the $b$-jet level ($\mu_{bb}$), ignoring that leptons are in fact observed and the invariant 
mass between the $b$-jet and the lepton ($M_{lb}$) which is very correlated 
with \mttp\ constructed with the $b$-jet+lepton combined.

In the analysis, the physics background is estimated using MC while the background with mistag
$b$-jets is evaluated using antitag events. The combinatoric background is suppresed using
a dedicated selection algorithm~\cite{cms:endpoint}.
The top-quark mass is extracted using a maximum likelihood fit of the endpoint of the three
chosen variables taking the object resolution into account.
Indeed in the limit of perfect object measurements, the maximum of the $\mu_{\ell\ell}$ distribution is 
equal to the $W$-boson mass (assuming zero neutrino mass), 
the maximum of the $\mu_{bb}$ distribution is equal to the top-quark mass
while the maximum of $M_{lb}$ can be expressed analytically using the energies and momenta of the daughter
of $t \to Wb$ in the top-quark rest frame. 
The fitted distributions are shown in Figure~\ref{fig:cmsendpoint}. 
Using this technique, CMS measures~\cite{cms:endpoint}:
$\mt = 173.9 \pm 0.9 {\rm (stat)} ^{+1.7}_{-2.1} {\rm (syst)}$~GeV.
The precision of this result is comparable to the one from the standard measurement in the same channel.
As can be seen in Table~\ref{tab:cmsendpoint}, the largest systematic uncertainty comes from
the uncertainty on the jet energy scale. 

\begin{figure}[!htb]
\centerline{
\includegraphics[width=0.7\textwidth]{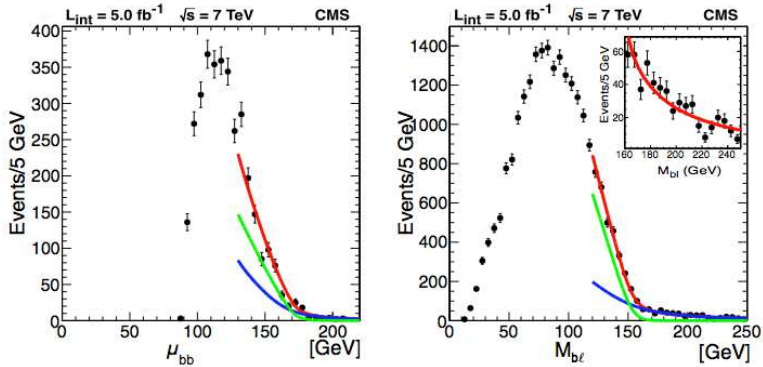}
}
\caption{Results for the endpoint fit in CMS where the red line represents the full fit while
the green and blue curves are for the signal and background shapes respectively~\cite{cms:endpoint}.}
\label{fig:cmsendpoint}
\end{figure}

\begin{table}
\centerline{
        \begin{tabular}{lc}
        \hline
        ~~~Source~~~    & ~~~ $\delta\mt$ (GeV)~~~ \cr\hline
        Jet Energy Scale & ${}^{+1.3}_{-1.8}$\cr
        Jet Energy Resolution &  ${\pm}0.5$\cr
        Lepton Energy Scale & ${}^{+0.3}_{-0.4}$\cr
        Fit Range &   ${\pm}0.6$\cr
        Background Shape &    ${\pm}0.5$\cr
        Jet and Lepton Efficiencies &     ${}^{+0.1}_{-0.2}$\cr
        Pileup &  $<$0.1 \cr
        QCD effects &     $\pm$0.6\cr\hline
        Total & ${}^{+1.7}_{-2.1}$\cr\hline
        \end{tabular}
}
\caption{Summary of the systematic uncertainties affecting the CMS measurement of the top-quark using
the endpoint method~\cite{cms:endpoint}.
}
\label{tab:cmsendpoint}
\end{table}

\section{Mass measurement using the B-hadron lifetime}
\label{sec:bhadron}

The top-quark mass can also be measured using different observables.
For instance the lifetime and decay length of the B-hadrons from the top-quark
decay depends almost linearly on the top-quark mass as can be seen in 
Figure~\ref{fig:lxymtop}. Alternatively the lepton \pt\ from the decay of the $W$-boson from the top
quark can also be used as a mass estimator.
The advantage of such estimators is that they minimally rely on the calorimeter-based
uncertainty like the jet energy scale uncertainty. However
these methods can potentially be rather sensitive to the modeling of the top production 
kinematics or to the calibration of the $b$ decay length or the $b$ fragmentation model.

\begin{figure}[!htb]
\centerline{
\includegraphics[width=0.4\textwidth,height=5cm]{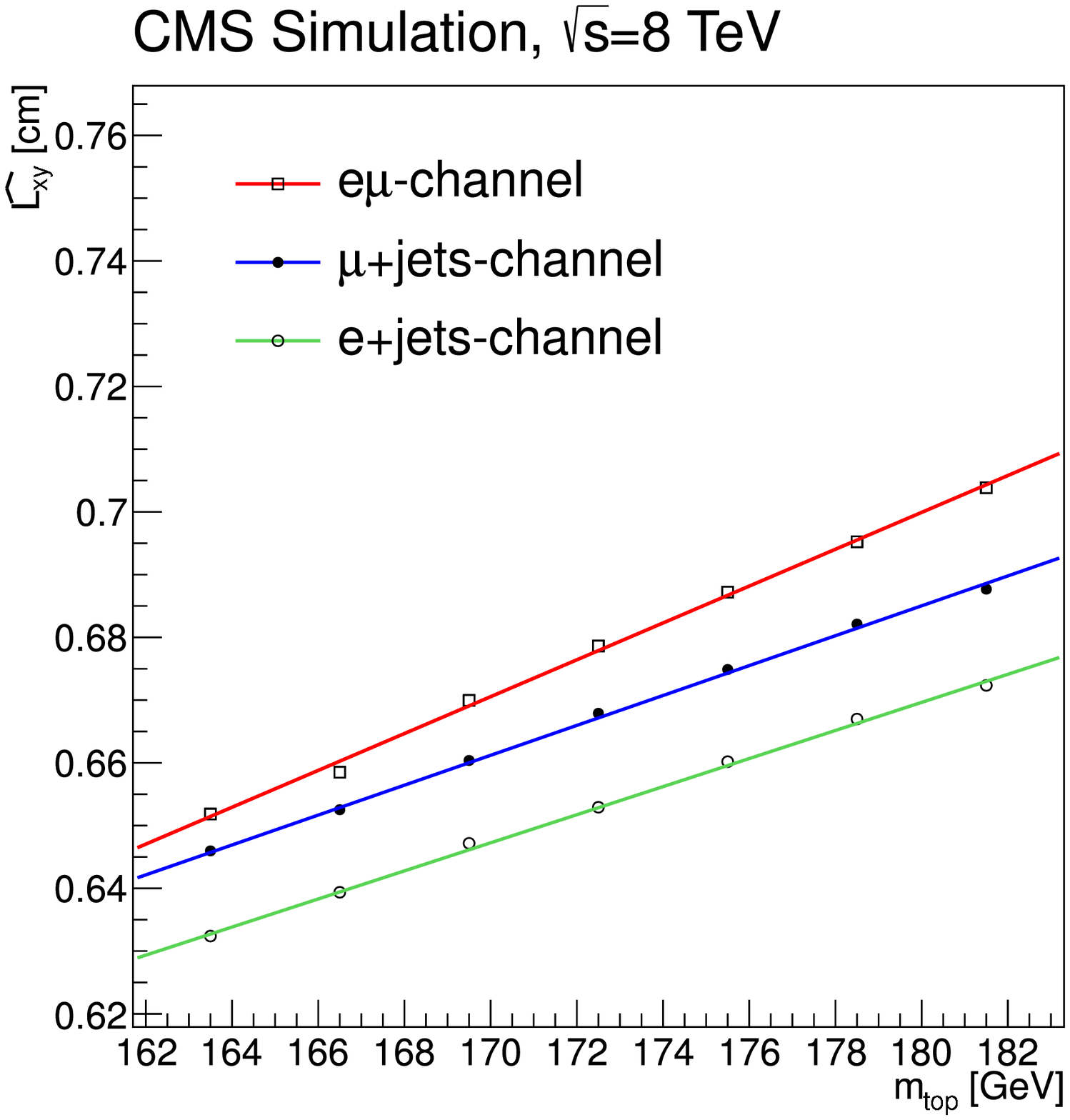}
}
\caption{Median of the transverse $b$ decay length distribution between the primary and the secondary 
vertex as a function of the simulated top-quark mass for three final states studied by CMS~\cite{cms:lxy}.}
\label{fig:lxymtop}
\end{figure}

\subsection{Measurement using the B-hadron lifetime at CDF}

These alternative methods were first developped at CDF in the \ljets\ channel with 
at least one $b$-tagged jet using 1.9~\ifb~\cite{cdf:lxy}.
The top-quark mass was simultaneously extracted from the B-hadron lifetime and from the
lepton \pt.
The main difficulty of this analysis appears to be the calibration of the transverse decay length.
Indeed corrections for the inaccuracy of the fragmentation simulation in EVTGEN has been necessary 
as well as corrections for the tracker modeling in the simulation.
These corrections are determined using a sample of $b\bar{b}$ events (with 95\% purity) 
as a function the \pt\ of jets reconstructed only in the tracker.
These track-based jets are previously calibrated using $\gamma$+jets events.
The uncertainty on the calibration of the transverse decay length are the dominant systematic 
uncertainty on the final result.
In the case of the measurement using the lepton \pt, the understanding of the lepton \pt\ scale
is the largest systematic uncertainty.
Constructing a combined likelihood shown in Figure~\ref{fig:cdflxy} with the two observables,
CDF measures~\cite{cdf:lxy}: $\mt = 170.7 \pm 6.3 {\rm (stat)} \pm 2.6 {\rm (syst)}$~GeV. 
Details on the systematic uncertainties limiting the measurements are presented in Table~\ref{tab:cdflxy}.

\begin{figure}[!hbt]
\centerline{
\includegraphics[width=0.5\textwidth]{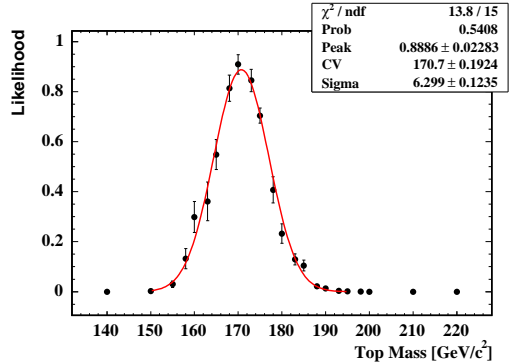}
}
\caption{Likelihood constructed from 23 \mt\ test points using the transverse B-hadron decay length 
and the lepton \pt\ in the top-quark decay by CDF~\cite{cdf:lxy}.}
\label{fig:cdflxy}
\end{figure}

\begin{table}
\centerline{
  \begin{tabular}{lccc}
  Systematic [$\textnormal{GeV}/c^2$]& Lxy & Lepton $p_T$ & Simultaneous\\
  \hline
  Background Shape & 1.0 & 2.3 & 1.7 \\
  QCD Radiation & 0.5 & 1.2 & 0.7 \\
  PDF & 0.3 & 0.6 & 0.5\\
  Generator & 0.7 & 0.9 & 0.3 \\
  Lepton $p_T$ Scale & 0 & 2.3 & 1.2 \\
  Lxy Calibration & 2.5 & 0 & 1.1 \\
  Multiple Interactions & 0.2 & 1.2 & 0.7 \\
  Calorimeter JES & 0.4 & 0.4 & 0.3\\
  \hline
  Systematics Total &  2.9 & 3.8 & 2.6 \\
  \end{tabular}
}
\caption{Final systematic uncertainties for the transverse B-hadron decay length 
and the lepton \pt\ CDF measurement~\cite{cdf:lxy}.
}
\label{tab:cdflxy}
\end{table}

\subsection{Measurement using the B-hadron lifetime at CMS}

CMS has adapted CDF method using both the \ljets\ and dilepton final state using
19~\ifb\ of LHC data at 8~TeV. 
In this analysis, the chosen observable is the median of the distribution of 
secondary vertices with maximal transverse decay length ($L_{xy}$).
The calibration for $L_{xy}$ is cross-checked using dijet events with one muon-tagged 
jet, taken to be the tag jet, while the second jet is taken to be the probe.
The distribution of the secondary vertex mass of this probe jet is then compared with
the prediction after fitting the light, $c$ and $b$-jets fractions.  
The agreement appears to be good as shown in Figure~\ref{fig:cmslxy}.

\begin{figure}[!hbt]
\centerline{
\includegraphics[width=0.4\textwidth]{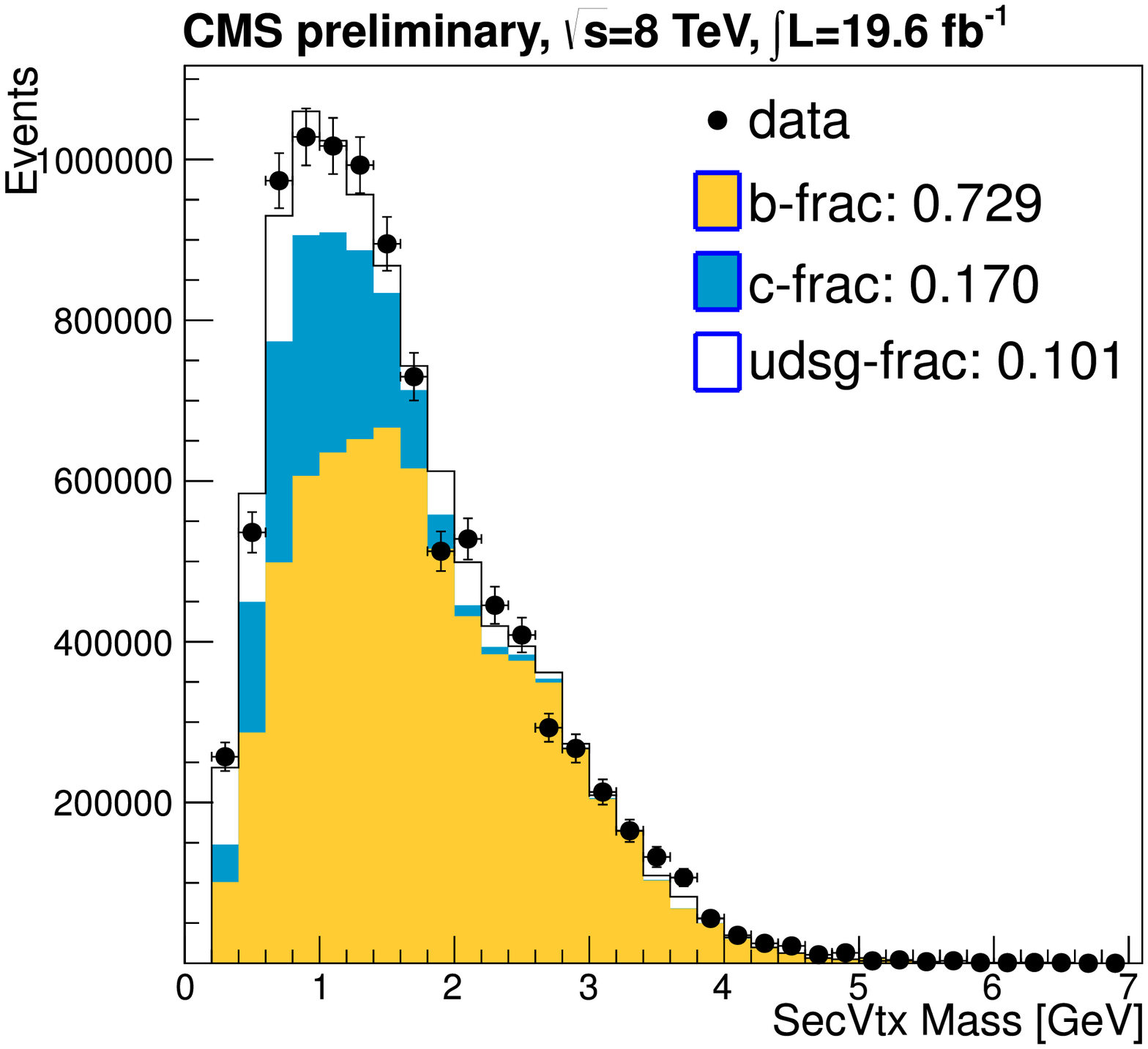}
}
\caption{Inclusive fit to the flavor content of a dijet sample based on the secondary vertex mass distribution 
to check the calibration of $L_{xy}$ in CMS B-hadron lifetime measurement~\cite{cms:lxy}.}
\label{fig:cmslxy}
\end{figure}

The top-quark mass extraction using the median values of $L_{xy}$ after calibration leads to~\cite{cms:lxy}:
$\mt = 173.5 \pm 1.5 {\rm (stat)} \pm 1.3 {\rm (syst)} \pm 2.6 {\rm (\pt(t))}$~GeV.
As can be seen in Table~\ref{tab:cmslxy}, the modeling of the top-quark \pt, which is mass dependent,
has a huge influence on the result. A systematic uncertainty based on reweighting the simulation to 
the unfolded top-quark \pt\ spectrum from data is assigned. This is currently the limiting uncertainty. 
In the future, the possibility to use an invariant quantity like the lepton-vertex invariant mass
could be studied since it would keep the information
on the top-quark mass while being less dependent on the top-quark kinematics.

\begin{table}
\centerline{
\begin{tabular}{llccc}
\hline
\hline
\multicolumn{2}{c}{Source}  & \multicolumn{3}{c}{$\Delta$\mt [GeV]} \\
& & $\mu$+jets& $e$+jets & $e\mu$\\\hline
\multicolumn{2}{l}{Statistical} & 1.0 & 1.0 & 2.0\\
\hline
\multirow{5}{*}{Experimental}
& Jet energy scale      &  $0.30\pm0.01$ & $0.30\pm0.01$ & $0.30\pm0.01$ \\
&Multijet normalization ($\ell$+jets)   & $0.50\pm0.01$& $0.67\pm0.01$ & - \\
&W+jets normalization ($\ell$+jets)    & $1.42\pm0.01$ & $1.33\pm0.01$ & - \\
&DY normalization ($\ell\ell$)           &-  & - & $0.38\pm0.06$\\
&Other backgrounds normalization   & $0.05\pm0.01$ & $0.05\pm0.01$ &
$0.15\pm0.07$ \\
&W+jets background shapes  ($\ell$+jets)& $0.40\pm 0.01$ & $0.20\pm
0.01 $ &- \\
&Single top background shapes & $0.20\pm 0.01$ & $0.20\pm 0.01$ & $0.30\pm0.06$\\
&DY background shapes  ($\ell\ell$) & - & - & $0.04\pm0.06$\\
& Calibration  & $0.42\pm0.01$ & $0.50\pm0.01$ & $0.21\pm0.01$\\
\hline
\multirow{4}{*}{Theory}
& $Q^{2}$-scale                     &  $0.47 \pm 0.13$ &  $0.20 \pm 0.03$ & $0.11\pm0.08$ \\
& ME-PS matching scale        &  $0.73 \pm 0.01$  & $0.87 \pm 0.03$ & $0.44\pm0.08$ \\
& PDF                                     & $0.26\pm0.15$ & $0.26\pm0.15$ & $0.26\pm0.15$\\
& Hadronization model         & $0.95\pm 0.13$ & $0.95 \pm 0.13$  & $0.67 \pm 0.10$\\
& B hadron composition       & $0.39\pm 0.01$ & $0.39\pm 0.01$ & $0.39\pm 0.01$\\
& B hadron lifetime            & $0.29 \pm 0.18$ & $0.29 \pm 0.18$ &  $0.29 \pm 0.18$  \\
& Top quark \pt modeling     & $3.27\pm0.48$ & $3.07\pm0.45$ & $2.36 \pm0.35$ \\
& Underlying event                & $0.27\pm0.51$  & $0.25\pm0.48$ & $0.19\pm0.37$ \\
& Colour reconnection           & $0.36\pm0.51$ & $0.34\pm0.48$ & $0.26\pm0.37$ \\
\hline
\hline
\end{tabular}
}
\caption{Statistical, experimental and theoretical systematic uncertainties on the measured 
top-quark mass based on the median of the transverse B-hadron decay length distribution
between the primary and the secondary vertex by CMS~\cite{cms:lxy}.
}
\label{tab:cmslxy}
\end{table}

\section{Conclusion}
\label{sec:conclusion}

Now that the precision on the direct top-quark mass measurements reaches 1~GeV, 
alternative methods that are less sensitive to MC
(and so less sensitive to the top-quark mass scheme implemented in MC) or 
with different sensitivity to systematic uncertainties need to be developped.
Some of these alternative approaches have been described here. 
For some of them the achieved precision is still modest.
However with plenty of statistics forseen,
the LHC Run 2 will enable to improve them allowing in particular to
study the systematic limitation using data.


\begin{footnotesize}

\end{footnotesize}


\end{document}